\documentclass[12pt]{article}
\newcommand{\citep}{\cite}
\usepackage{amsfonts}
\usepackage{amssymb}
\usepackage{epsfig}
\usepackage{rotating}
\usepackage{lineno}
\usepackage{color}
\usepackage{amsmath}
\usepackage{mathrsfs}
\usepackage{hyperref}
\usepackage{setspace}
\usepackage{grffile}
\usepackage{multirow}
\usepackage{verbatim}   
\usepackage{url}
\usepackage{bm}
\usepackage{amsmath}
\usepackage{graphicx}
\usepackage{multirow}
\newcommand{\avg}[1]{\left\langle#1\right\rangle}

\newcommand{\Reff}{\mathcal{R}_\text{eff} }
\newcommand{\R}{\mathcal{R}}

\begin{document}

\title{The Effect of Reactive Social Distancing on the 1918 Influenza Pandemic}
\author{
 Duo Yu$^{1,2,}$\footnote{
These authors equally contributed to this work.}, Qianying Lin$^{1,*}$, Alice PY Chiu$^{1}$\footnote{alice.py.chiu@polyu.edu.hk} and Daihai He $^{1}\footnote{daihai.he@polyu.edu.hk}$
\vspace{0.2cm}\\
{\footnotesize $^1$ Department of Applied Mathematics, Hong Kong Polytechnic University}\\
{\footnotesize Hung Hom, Kowloon, Hong Kong (SAR) China}\\
{\footnotesize $^2$ Department of Biostatistics, School of Public Health, University of Texas}\\
{\footnotesize Health Science Center at Houston, United States}\\
}
\maketitle

\begin{abstract}
The 1918 influenza pandemic was characterized by multiple epidemic
waves. We investigated into reactive social distancing, a form of
behavioral responses, and its effect on the multiple influenza waves
in the United Kingdom. Two forms of reactive social distancing have
been used in previous studies: Power function, which is a function
of the proportion of recent influenza mortality in a population,
and Hill function, which is a function of the actual number of
recent influenza mortality.  Using a simple epidemic model with a
Power function and one common set of parameters, we provided
a good model fit for the observed multiple epidemic waves in London
boroughs, Birmingham and Liverpool. Our approach is different from previous studies where separate models are fitted to each city. We then applied these model parameters obtained from fitting three cities to all 334 administrative units in England and Wales and including the population sizes of individual administrative units. We computed the Pearson's correlation between the observed and simulated data for each administrative unit. We achieved a median correlation of 0.636, indicating our model predictions perform reasonably well. Our modelling approach which requires reduced number of parameters resulted in computational efficiency gain without over-fitting the model. Our works have both scientific and public health significance.
\end{abstract}

\section{Introduction}
The influenza pandemic of 1918 had been regarded as the deadliest
pandemic in recorded history, as it had caused 50 million deaths
worldwide \cite{Johnson2002}. Due to its exceptional lethality and
unusual epidemiological features, an in-depth understanding of the
1918 pandemic can provide insights to future influenza pandemic control
planning. The 1918 pandemic was characterized by multiple waves of
mortality. In the United Kingdom, the pandemic took place as three
distinct waves: the first wave in the summer of 1918, the second
wave in autumn of the same year and the third wave in the spring
of 1919. Earlier studies attempted to identify the underlying
causes of multiple waves
\cite{Bootsma2007,Daihai2011,Daihai2013,Caley2008}. They all pointed
towards human behavioral responses being a key factor in the
exhibition of multiple waves in the 1918 influenza pandemic.  Human
behavioral responses had been considered as a crucial factor that
could influence the transmission of infectious diseases
\cite{Ferguson2007,Funk2010,Valle2005,Epstein2008,Funk2009,TC2010,Poletti2009}.
When an infectious disease invaded a population, people behaved reactively to reduce the probability of being infected, such as avoidance of
mass gathering, putting on face masks and actively maintaining
personal hygiene.  He et al. showed that the temporal changes in
transmission rates could be the most likely explanation for the three
epidemic waves in England and Wales, and human behavioral changes
had the largest effect on the epidemic waves of weekly cases \cite{Daihai2011,Daihai2013}. Poletti et al. studied the 2009 H1N1
influenza pandemic and concluded that human behavioral changes could
have a significant impact on the timing, dynamics and magnitude of
the epidemic spread \cite{Poletti2009}.  Bootsma and Ferguson found
that the effect of reactive social distancing could have a stronger impact
on the epidemic waves of the weekly cases than on the overall mortality
\cite{Bootsma2007}. However, the impact of reactive social distancing
on the final epidemic size remained unknown. In previous studies,
an epidemic model with a common set of parameters for different
cities did not fit the observed data well \cite{Bootsma2007}. Here,
we compare two mathematical functions of reactive social distancing:
Power function, which is a function of the proportion of recent
influenza mortality in a population, and Hill function, which is a
function of the actual number of recent influenza mortality. This manuscript is arranged as follows: We fitted out model with the same set of input parameters to the observed data in London boroughs, Birmingham and Liverpool. We applied these model parameter values obtained from fitting three cities to all 334 administrative units in England and Wales and including the population sizes of individual administrative units. We demonstrated the impact of reactive social distancing on the final epidemic size. In the appendix, we showed the theoretical results of the oscillations induced by reactive social distancing.

\section{Methods}
\subsection{Data}
We analyzed data on weekly influenza deaths (either influenza was recorded as contributing or
death has been assigned to influenza) between June 29, 1918
and May 10, 1919 from 334 Administrative units in England for London
boroughs, Birmingham and Liverpool based on the publicly available
data \cite{Johnson2001}. We obtained the daily temperatures from
the UK Met Office (http://www.metoffice.gov.uk).

\subsection{Model}

\begin{figure}[!h]
\centering
\includegraphics[width=12cm]{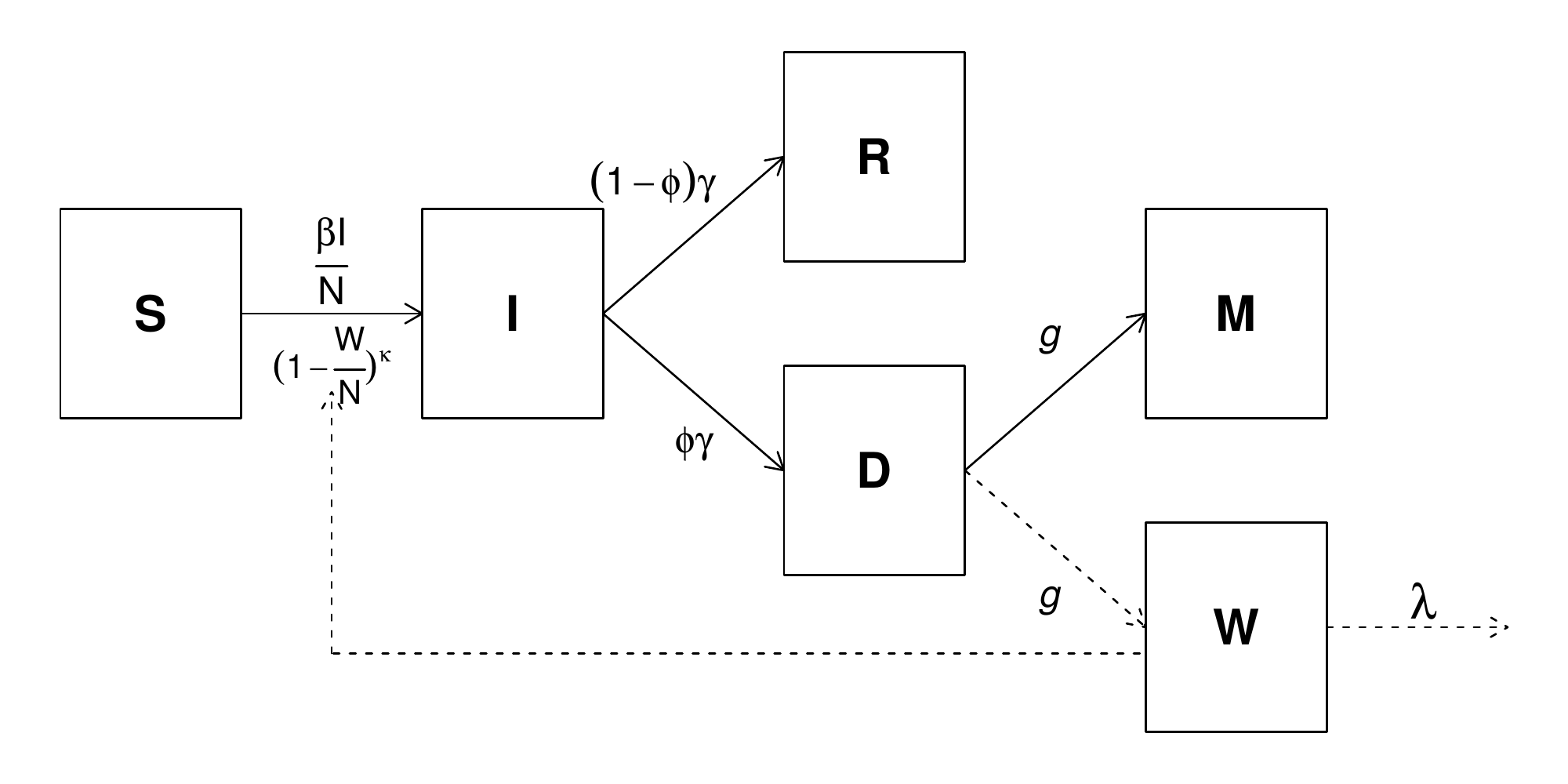}
\caption{Schematic diagram shows the transmission dynamics during a disease outbreak.
$S$ denotes the number of susceptible individuals, $I$ denotes the number of infectious individuals,
$R$ denotes the number of recovered individuals, $D$ denotes the number of individuals who are no longer infectious and are progressing to death of influenza or pneumonia causes, $M$ denotes the cumulative number of influenza-related deaths, and $W$ denotes recent influenza mortality.
}
\label{model}
\end{figure}

Bootsma and Ferguson \cite{Bootsma2007} proposed that people reduce
their exposure to potentially infectious contacts in response to
high mortality rates during an influenza pandemic. In this study,
we assumed the perceived risk of influenza is proportional to the
number of recent deaths, and that people practised social distancing
in response to this perceived risk. Here, we employed a simple
Susceptible-Infectious-Recovered (SIR) model to account for such
behavioral changes in the population \cite{Daihai2013}. Our earlier study
shows that similar results are achieved if an additional ``exposed" class is included
\cite{Daihai2013}. The model is represented as follows:

\begin{equation}\label{model1}
\begin{split}
&\dot{S}=-\frac{\beta(t)}{N}SI,\\
&\dot{I}=\frac{\beta(t)}{N}SI-\gamma I,\\
&\dot{R}=\gamma (1-\phi) I,\\
&\dot{D}=\gamma \phi I-g D,\\
&\dot{M}=g D, \\
&\dot{W}=g D-\lambda W.
\end{split}
\end{equation}
Here, $S$ denotes the number of susceptible individuals, $I$
denotes the number of infectious individuals, $R$ denotes the number of individuals who are immune to the disease. $D$ denotes the number of individuals who are no longer infectious and are progressing to deaths of influenza or pneumonia causes.
$M$ denotes the number of influenza deaths. $W$ denotes the recent influenza mortality, and we assume that the general public's risk perception is based on W.
$N$ is the population size which is assumed to be constant. The population sizes for London Borough, Birmingham and Liverpool were approximately 4,484,523, 919,444 and 802,940 during the study period, respectively. Parameter $\phi$ stands for case-fatality-ratio (CFR).

Parameters $\gamma$, $g$ and $\lambda$ are rates at which individuals
move from one class to the next class. $\gamma^{-1}$ and $g^{-1}$
are the mean infectious period (to  be fixed at 4 days \cite{Mills2004})
and the mean time from loss-of-infectiousness to death (to be fixed
at 8 days). Thus the mean duration from infection to death is 12
days \cite{Bootsma2007}. $\lambda^{-1}$ is the duration of human
risk-reduction behavior in days. Following \cite{Daihai2013},
$\beta(t)$ is the transmission rate and takes the following form:
\begin{equation} \begin{split} \beta(t) &=\beta_{0}\cdot e^{-\xi
T(t)}\cdot [1+\alpha H(t)] \cdot \left[1-\frac{W(t)}{N}\right]^{\kappa}
\end{split} \end{equation}
where there are four components

(i) $\beta_0$ is the constant
baseline transmission rate.

(ii) $e^{-\xi T(t)}$ is the term representing the temperature effect, $T(t)$ is time series of daily temperature (the
model is simulated with a step size of 1 day), the parameter $\xi$
controls the intensity of the temperature effect.

(iii) $1+\alpha H(t)$ is the factor of school terms with amplitude parameter $\alpha$
and school day function $H(t)$ (a step function takes a large value
on school days and a small value otherwise, see \cite{Daihai2013});
Easter and Christmas holiday dates are known. However, the summer vacation period ($t_1$, $t_2$) is unknown, therefore the start
date ($t_1$) and end date ($t_2$) are to be estimated.
During the 1910's, a large number of people, including both adults and school children, are involved in summer harvesting. Thus there is a substantial impact on influenza transmission. This was discussed in \cite{Daihai2013}).

(iv) The last factor is the human behavioral term, where $W(t)$ denotes the recent influenza mortality, and $\kappa$
represents the intensity of human behavioral response in relation to the risk of influenza infection.

Finally, we define the basic reproductive number $\R_0=\frac{\avg{\beta(t)}}{\gamma}$ with $W(t)=0$ (see \cite{Daihai2013}, we have $W=0$ at disease-free equilibrium). Note that $\avg{\beta(t)} < \beta_0$, since  the temperature effect has an average value of smaller than 1. However,
with human behavioral reaction, it is more effective to estimate the effective reproductive number, $\Reff(t)$ instead, which is defined as the actual average number of secondary cases per primary case of infection at time $t$ \cite{ChowNish}. We have
$\Reff(t)=\beta(t)S(t)/(\gamma N)$\cite{Chowell2004}. If $\Reff(t)\leq 1$, it indicates that the epidemic is under control.

Here, we use the formula $[1-\frac{W(t)}{N}]^{\kappa}$ to represent
the effect of reactive social distancing on transmission rate.
Bootsma and Ferguson \cite{Bootsma2007} modelled the reactive social
distancing as a Hill function  $\frac{\kappa}{\kappa+W(t)}$.
If we compare the Taylor's series expansion in the Power function and the Hill function, we noted that $N/\kappa$ in the Hill function is equivalent to the $\kappa$ in the Power function. Furthermore, if we replace $\kappa$ with $N/\kappa$ in the Hill function, we will obtain a modified-Hill function. The first two terms in the Taylor's series expansion in both the Power function and the modified-Hill function are almost the same.

\begin{subequations}\label{te}
\begin{eqnarray}
\mbox{Power}& \left[1-\frac{W(t)}{N}\right]^{\kappa}&=1-\frac{\kappa }{N}W(t)+\frac{1}{2}\frac{\kappa(\kappa-1)}{N^{2}}W(t)^{2}+\cdots\\
\mbox{Hill} & \frac{\kappa}{\kappa+W(t)}&=1-\frac{1}{\kappa} W(t)+ \frac{1}{\kappa^{2}}W(t)^{2}+\cdots \\
\mbox{Modified Hill} &\frac{N/\kappa}{N/\kappa+W(t)}&=1-\frac{\kappa }{N}W(t)+ \frac{\kappa^{2}}{N^{2}}W(t)^{2}+\cdots
\end{eqnarray}
\end{subequations}
The above suggests that the Power function and the modified-Hill function
will lead to almost the same results. The key point is whether $W(t)$ is scaled by $N$ or not in the
behavioral term.

\subsection{Model Fitting and Parameter estimates}
We fit the model as described in Figure ~\ref{model} to the reported
weekly influenza deaths from the three largest administrative units:
London boroughs, Birmingham and Liverpool during the 1918-1919
influenza pandemic. We assumed the epidemiological parameters remained
the same for all three cities, and then further modelled for all
334 administrative units in England and Wales by incorporating their
respective population sizes $N$. Previous studies
\cite{Bootsma2007,Daihai2013} assumed that key parameter values being
distinct for different cities or administrative units. Thus we
greatly reduce the number of free parameters and computational time
in this work. Our assumption is also biologically plausible as
previous studies showed that the transmissibility during the pandemic
showed little spatial variations \cite{Chowell2008, Eggo2011}.

Partially observed Markov process (POMP) model within a plug-and-play
framework \cite{Daihai2010} was applied to simulate the epidemic
dynamics in equation~\ref{model1}. Parameters including baseline
transmission rate ($\beta_{0}$), case-fatality-ratio ($\phi$),
school-term impact intensity ($\alpha$), air temperature impact
intensity ($\xi$), reactive social distancing intensity ($\kappa$)
and rate of recovery of social distancing ($\lambda$) are estimated
by the Iterated Filtering method, allowing us to compute the maximum
likelihood \cite{Ioni2006,Ioni2011}.

Using this method, stochastic perturbation is added to the unknown parameters
for the exploration of parameter space. This allows us to extend the range of global search, to avoid local minima and to construct an approximation to derive the log-likelihood. Selection of the estimates is achieved by Sequential Monte
Carlo(SMC) or particle filtering, which keeps the results to be consistent with the data. With well-constructed procedures and continually decreasing perturbations, the iterations will converge to the maximum likelihood estimates. This method has been widely used in infectious disease modelling studies, including Ebola, cholera, malaria, influenza, as well as studies in finance and ecological dynamics.
The 'pomp' package in R is used for implementation (\url{http://kingaa.github.io/pomp/}).

\section{Results}

\begin{figure}[h!]
\centering\includegraphics[width=18cm]{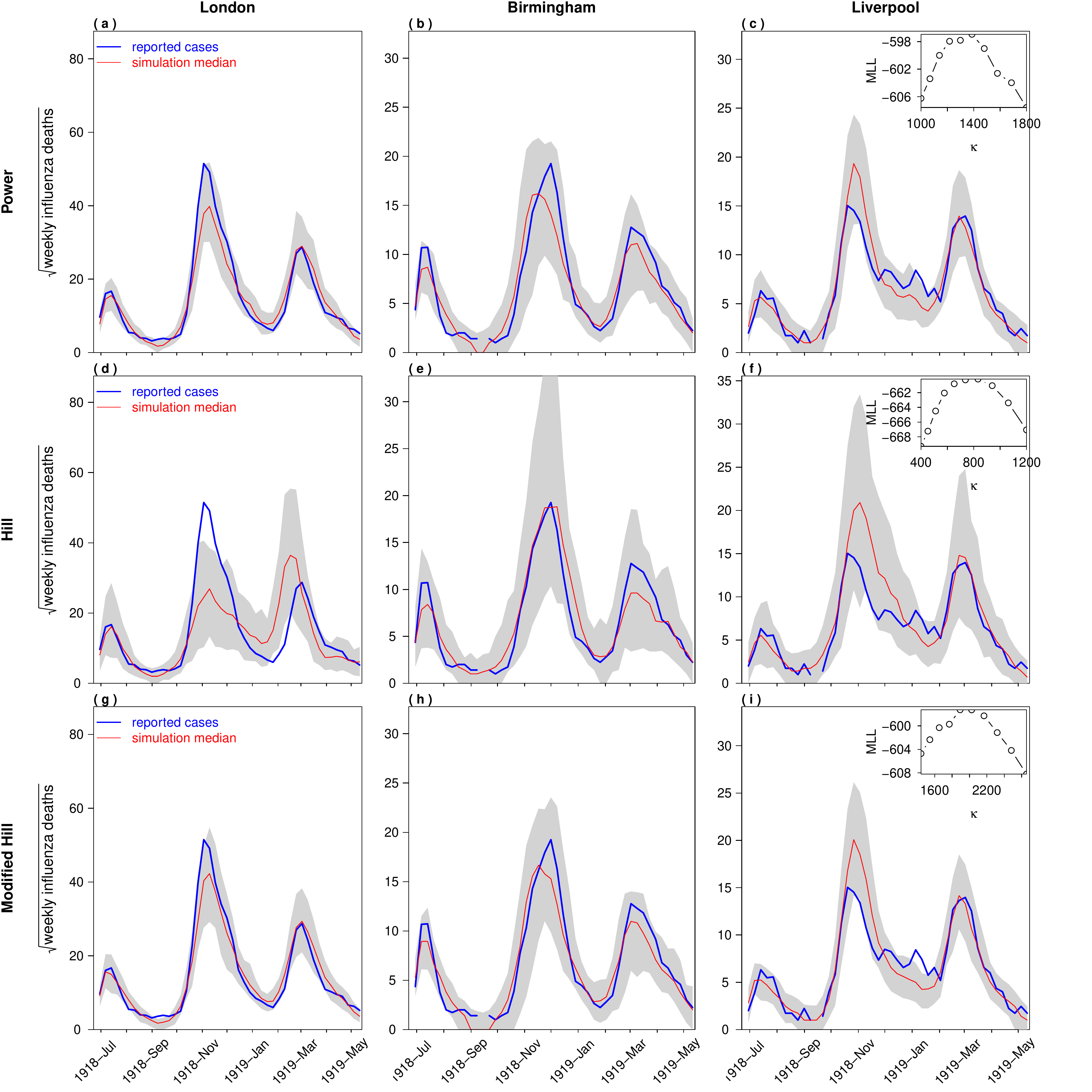}
\caption{Simulation comparisons of the weekly influenza deaths
during the 1918 pandemic in London, Birmingham and Liverpool. We
used human behavioral terms of the (a-c) Power function, (d-f) Hill
function, and (g-i) modified-Hill function. Blue bold line depicts the
reported cases. Red thin line depicts the simulation median. The shaded
area indicates the 95\% confidence interval. The inset panels show the profile log-likelihood
as a function of $\kappa$.} \label{fitting}
\end{figure}

Figure~\ref{fitting} shows the best-fitting simulation models using
three different functions, i.e. a model with the Power function
(panels a-c), the Hill function (panels d-f) and the modified-Hill
function (panels h-j).  The inset panels show the log-likelihood
profile of each model as a function of the parameter $\kappa$. Since
the numbers of parameters of the three models (i.e. their model
complexity) are the same, we can directly compare their Maximum Log
Likelihoods (MLL). The MLLs for the Power function, the Hill function
and the modified-Hill function are -596.12, -659.58 and -596.34
respectively. Since a larger MLL indicates a better model fit, we
conclude that both the Power function and the modified-Hill function provide the best model choice.
Thus, we identify a unified model (with the Power function or the modified-Hill function) for the three waves in these three
cities. The Power function and the modified-Hill function virtually achieve the same goodness-of-fit levels, and both are superior to the Hill function used in \cite{Bootsma2007}, using our current dataset.

In previous studies \cite{Bootsma2007,Daihai2013}, key
parameters such as $\R_0$ and $\kappa$ are assumed to be different for each city in order to achieve the best model fit.
Our model fit as shown in Figure~\ref{fitting}(a-c) are reasonably well, which use a unified model
that estimated the same set of epidemiological parameters (i.e., the same $\kappa$ in all three cities), but using different population sizes and initial conditions for each of the three cities.

\begin{figure}[h!]
\centering
\includegraphics[width=8cm]{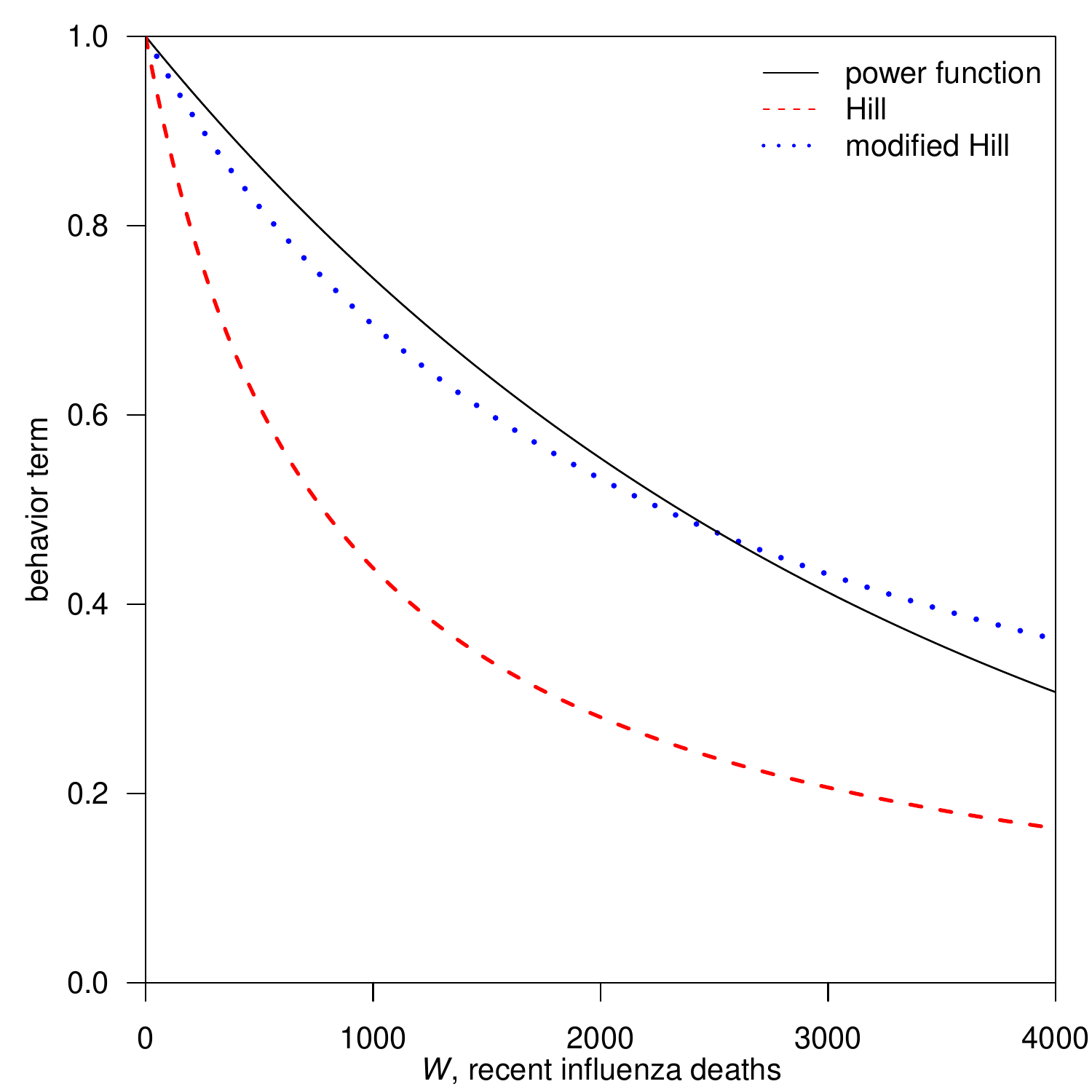}
\caption{Comparison among the value of three behavioral functions with their best-fitted $\kappa$.
} \label{term}
\end{figure}

Figure~\ref{term} compares the three behavioral functions with the maximum likelihood estimates of $\kappa$. We can see that the Power function and the modified-Hill function largely overlaps, while the Hill function is clearly different from the other two functions.

\begin{figure}[h!]
\centering
\includegraphics[width=18cm]{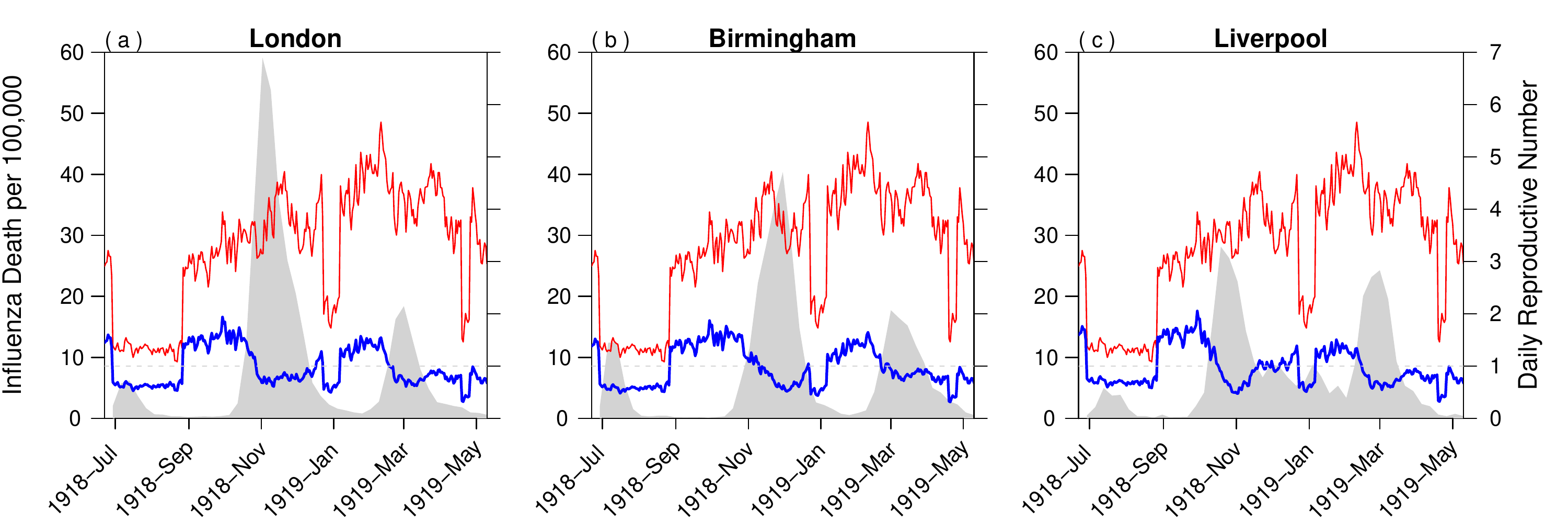}
\caption{Estimated daily basic reproductive number $\R_0$ (red
thin curve), effective reproductive number $\Reff$ (blue bold curve) and weekly
influenza deaths (shaded region). The average basic reproductive number is 3.24 in the three cities.}
\label{transmission}
\end{figure}

Figure~\ref{transmission} shows the estimated daily reproductive
number $\R_0$ (red thin curve) and the effective reproductive number $\Reff$ (blue bold curve).
The estimated daily reproductive number is identical in the three cities, since we assume all related parameters
to be identical. The fluctuations in the daily reproductive number are due to school term and daily temperature, whereas
we set $W=0$ (see \cite{Daihai2013}). In the effective reproductive number, we use the estimated $W(t)$ and susceptible $S(t)$.
Thus $\Reff$ is distinct among the cities and when it is greater than 1 (or less than 1), the mortality curve increases (or decreases) with a time delay of about 12 days.


Table \ref{summary} summarizes all parameters used in the model
with the Power function. All parameters values are largely biologically
reasonable \cite{Bootsma2007,Daihai2013}.
Thus, we successfully find a model with common parameter values
for all three cities. The only differences are in the population
sizes and the initial conditions. We denote the initial susceptible
population and initial population size to be $S_{0}$ and $I_{0}$,
respectively. Note that the estimated initial conditions are similar
among the three cities.

\begin{table}[h!]
\begin{center}
\caption{Summary of all parameters estimated in the best-fitted model with a Power function.}\label{summary}
\begin{tabular}{c|ccc|c}
\hline
   Parameter     & London boroughs & Birmingham & Liverpool & Type \\
\hline
initial, $S_0/N$ & 0.685 (0.503, 0.836) & 0.677 (0.450,0.806) & 0.632 (0.253, 0.950) & distinct \\
initial, $I_0$ & 9552 (7116, 14611) & 3760 (2302, 4964) & 927 (610, 1521) & distinct\\
behavioral, $\kappa$ & \multicolumn{3}{c|}{1323.2 (1185.9, 1484.8)} & common\\
delay, $\lambda^{-1}$ (days)& \multicolumn{3}{c|}{12.43 (10.73, 14.55)} & common \\
CFR, $\phi$  & \multicolumn{3}{c|}{0.0118 (0.0108, 0.0129)} & common \\
baseline, $\beta_0$ & \multicolumn{3}{c|}{4153.8 (3067.9, 8424.1} & common \\
school-term, $\alpha$  & \multicolumn{3}{c|}{0.437 (0.377, 0.498)} & common \\
temperature, $\xi$ & \multicolumn{3}{c|}{0.04048 (0.03441, 0.04568)} & common\\
summer vacation  start & \multicolumn{3}{c|}{June 23 (May 24, June 28)} & common\\
summer vacation  end & \multicolumn{3}{c|}{August 21 (Aug 12, August 31)} & common \\
\hline
\end{tabular}
\end{center}
\end{table}

\subsection{Applying the model to 334 administrative units}
\begin{figure}[h!]
\centering
\includegraphics[width=16cm]{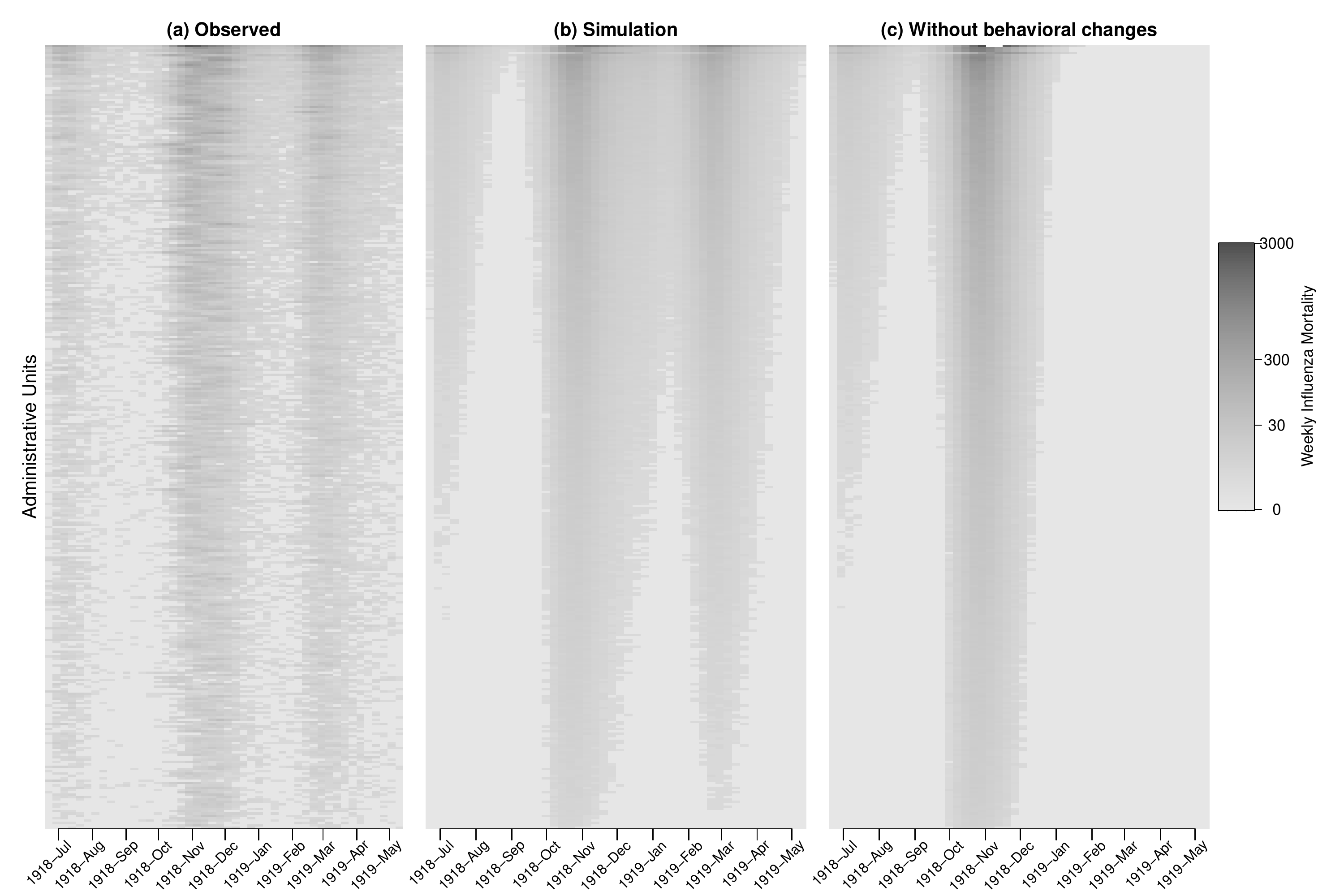}
\caption{Comparison between the observed and simulated patterns of influenza deaths in 334 administrative units. (a) Observed. (b) Simulation with all three factors (school term, temperature, and behavioral changes). (c) Without behavioral changes. Administrative units are ordered in descending population sizes from top to bottom.}
\label{admin}
\end{figure}

We apply our model with parameter values from fitting the three
largest cities to all 334 administrative units in England and Wales.
The only parameter we need to incorporate into this model is the
population size of each administrative unit. As the three cities
had similar initial conditions, we use that of Liverpool's. We
display our results in Figure~\ref{admin}. We compute the Pearson's correlations between the observed (panel (a)) and simulated values (panel (b), with all three factors) for each administrative unit. The median correlation is 0.636. This shows that our model performs reasonably well in at least half of the 334 administrative units while only using data from three major cities. If the effects of behavioral changes are removed, previous studies \cite{Bootsma2007,Daihai2013} show that the model cannot fit the data well even when the cities are being fitted separately. In panel (c), we show the simulation results of the 334 administrative units by setting $W=0$ and using other parameters from panel (b). It can been seen that model can only yield two wave, with the winter wave mmissed. Furthermore, we compare the overall attack rates in the two scenarios: Using all three factors, the estimated infection attack rate is about 28.5\% (95\% confidence interval: 14.1\%, 35.9\%). Without behavioral changes, the estimated infection attack rate is about 40.8\% (95\% CI: 34.0\%, 46.9\%). Thus the reduction due to behavioral changes is substantial.

\section*{Discussion and Conclusions}

Through a simple epidemic model with a Power function as behavioral
term and a common set of parameters, we provided a good model fit
for the observed multiple waves of influenza deaths in London
boroughs, Birmingham and Liverpool. Our results are novel compared
with earlier works by Bootsma and Ferguson \cite{Bootsma2007} and
He et al \cite{Daihai2013}. We also showed that the same set of
parameters can be used for modelling the 334 administrative units.
Furthermore, through given parameter values of mean infectious
period and mean time from loss-of-infectiousness-to-death, we showed
that there is an almost perfect linear relationship between the
mean period of damping oscillation and the duration of reactive
social distancing. Our theoretical damping oscillation results
provided a plausible explanation to the observed multiple waves,
where by reactively responding to the high proportion of influenza
deaths with social distancing, the epidemic waves will be dampened.
However, with the decline in the proportion of influenza deaths,
public risk perception will be lowered as well, and the reduced
social distancing could induce another epidemic wave. We also showed
that reactive social distancing leads to a reduction in final
epidemic size.

Our findings are plausible and are consistent with earlier mathematical
modelling studies on the 1918 influenza pandemic \cite{Mills2004, Ferguson2005}.
Bootsma and Ferguson \cite{Bootsma2007} developed an epidemic model to study
the impacts of public health interventions on the 1918 influenza
pandemic in 16 U.S. cities. He et al \cite{Daihai2013} proposed
another epidemic model which incorporates school opening and closing,
temperature changes and changes in human behavioral response during
the 1918 influenza pandemic in 334 administrative units of England
and Wales. However, in both of these studies, instead of using a
common set of model input parameters, unique model input parameters
were needed for model fitting of each city or administrative units.
Here, our model requires only a common set of parameters for the
three-city or the subsequent 334-administrative model-fitting
procedure, and the reduced number of model input parameters used represented significant improvement in computational
efficiency and resulted in more robust estimates. Caley et al
\cite{Caley2008} studied the 1918 influenza pandemic in Australia,
showing that reactive social distancing had a significant impact
on the observed multiple epidemic waves and final epidemic size.
Our effective reproductive numbers are comparable to these studies.

Compared to previous studies, our methods provide several improvements.
Our study has the following strengths. First, by using a common set
of model parameters for fitting the three-city model, we have greatly
enhanced our computational efficiencies and have also resulted in
more robust estimates of the final epidemic size. Second, we have
identified an almost perfect linear relationship between the mean
period of damping oscillations and duration of reactive social
distancing on various combinations of mean infectious period and
mean time from loss-of-infectiousness-to-death, which will be useful
for future studies on the time between influenza epidemic waves.
Major limitations of our study include the lack of direct historical
behavioral data on quantifying the extent of reactive social
distancing. Also, other non-pharmaceutical interventions could have
played a role on the influenza pandemic patterns observed, but these
measures are not considered in our model. There could be differences in summer vacation periods and daily temperature data in the three cities. However, such detailed data are not accessible to us. We could only make a simplifying assumption that they are the same across all locations.
In future epidemics or pandemics, such information will be available and could be incorporated into the framework developed in this work.

In conclusion, a simple model with reactive social distancing,
weather conditions, and school term could explain the observed
multiple waves and final epidemic size in London boroughs, Birmingham
and Liverpool during the 1918 influenza pandemic. Despite societal
changes, our historical analyses on the 1918 pandemic could still
serve as an important reference for future pandemic planning.

\section*{Acknowledgment}
We were supported by General Research Fund (Early Career Scheme) from Hong Kong Research Grants Council (PolyU 251001/14M) and
Start-up Fund for New Recruits from Hong Kong Polytechnic University.

\newpage

\end{document}